\documentstyle[sprocl]{article}

\input{psfig}
\def\be{\begin{equation}}
\def\ee{\end{equation}}
\def\bea{\begin{eqnarray}}
\def\eea{\end{eqnarray}}
\newcommand{\lsim}{\mathrel{\mathop{\kern 0pt \rlap
  {\raise.2ex\hbox{$<$}}}
  \lower.9ex\hbox{\kern-.190em $\sim$}}}
\newcommand{\gsim}{\mathrel{\mathop{\kern 0pt \rlap
  {\raise.2ex\hbox{$>$}}}
  \lower.9ex\hbox{\kern-.190em $\sim$}}}
\newcommand{\gagamma}{g_{a\gamma\gamma}}
\begin{document}

\title{THEORETICAL EXPECTATIONS AND\\
       EXPERIMENTAL PROSPECTS FOR SOLAR AXIONS  \\
       SEARCHES WITH CRYSTAL DETECTORS}
%SIGNALS FOR AXION SEARCHES IN CRYSTAL DETECTORS:\\
%       PRESENT SITUATION AND FUTURE PROSPECTS
%Solar Axion Searches with Crystalin Canfranc with a Ge Detector \\
%                  and prospects with other crystals}

\author{S. SCOPEL,
        I. G. IRASTORZA, S. CEBRI\'{A}N, E. GARC\'IA,
         D. GONZ\'{A}LEZ, \\A. MORALES, J. MORALES, A. ORTIZ de
         SOL\'{O}RZANO,\\
         J. PUIMED\'{O}N, A. SALINAS, M. L. SARSA, J. A. VILLAR}

\address{Laboratorio de F\'{\i}sica Nuclear. Universidad de Zaragoza\\
              50009, Zaragoza, SPAIN}

\maketitle\abstract{
A calculation of the expected signal due to Primakov coherent conversion of
solar axions into photons via Bragg scattering in several solid--state detectors is
presented and compared with present and future experimental sensitivities.
The axion window $m_a\gsim 0.03$ eV (not accessible at present by
other techniques)
could be explored in the foreseeable future
with crystal detectors to constrain the axion--photon coupling
constant $\gagamma$ below the latest bounds coming from
helioseismology. On the contrary
a positive signal in the sensitivity region of such devices would imply revisiting
other more stringent astrophysical limits derived for the same range of the axion mass.
}
%. The potentialities
%of this technique is critically discussed in light of the
%most recent developments in the field and compared to the present
%and future experimental sensitivities.
%An experiment currently under way in
%the Canfranc laboratory is briefly described.

\section{Introduction}
Introduced twenty years ago as the Nambu--Goldstone boson of the Peccei--Quinn symmetry
to explain in an elegant way CP conservation in QCD \cite{PQ}, the axion
is remarkably also one of the best candidates
%, along with supersymmetric
%particles,
to provide at least a fraction of the Dark Matter needed in Cosmology
in order to explain both gravitational measurements and models of structure formation.

Axion phenomenology is determined by its mass $m_a$
which in turn is fixed by the scale $f_a$ of the Peccei--Quinn symmetry
breaking, $m_a$ $\simeq$ 0.62 eV $(10^7$ ${\rm GeV}/f_a)$.
No hint is provided by theory about where the $f_a$ scale should
be.
% spanning in principle over several orders of magnitude.
A combination of astrophysical and nuclear physics
constraints, and the requirement that the axion relic abundance
does not overclose the Universe, restricts the allowed range of
viable axion masses into a relatively narrow window:
\begin{eqnarray}
 10^{-6}  {\rm eV}  \lsim & m_a & \lsim 10^{-3} {\rm eV} \nonumber \\
3  \;{\rm eV} \lsim  & m_a &  \lsim  20\;  {\rm eV}.
\label{eq:limits}
\end{eqnarray}
%A great experimental effort is presently under way in order to
%explore this range of masses.
The physical process used in axion search experiments
is the Primakov effect. It makes use of the coupling between the
axion field $\psi_a$ and the electromagnetic tensor:
\begin{equation}
{\cal L}=g_{a\gamma \gamma}\psi_a\epsilon_{\mu \nu \alpha
\beta}F^{\mu \nu} F^{\alpha \beta}=
\gagamma \psi_a \vec{B}\cdot\vec{E}
\label{eq:lagrangiana}
\end{equation}
\noindent
and allows for the conversion of the axion into a photon.

Solid state detectors provide a simple mechanism for axion detection
\cite{zioutas,creswick}.
Axions can pass in the
proximity of the atomic nuclei of the crystal where the intense electric field can
trigger their conversion into photons.
In the process the energy of the outgoing photon is equal to that of the
incoming axion.

Axions can be efficiently produced in the interior of the Sun
by Primakov conversion of the blackbody photons in the fluctuating electric field
of the plasma.
The resulting flux has an outgoing average axion energy
$E_a$ of about 4 keV (corresponding to
the temperature in the core of the Sun, $T\sim 10^7 K$)
that can produce detectable x--rays in a crystal detector.
Depending on the direction of the incoming axion flux with
respect to the planes of the crystal lattice, a coherent effect can be produced
when the Bragg condition is fulfilled,
leading so to a strong enhancement of the
signal. A correlation of the expected count--rate
with the position of the Sun in the sky is a distinctive signature of the axion
which can be used, at the least, to improve the signal/background
ratio.

The process described above is independent on
$m_a$ and so are the achievable bounds
for the axion--photon coupling
$\gagamma$.
This fact is particularly appealing, since
other experimental techniques
are limited to a more restricted mass range:
``haloscopes''\cite{cavities}, that use electromagnetic cavities
to look for the resonant conversion
into microwaves of non relativistic cosmological dark halo axions,
do not extend their search beyond $m_a\simeq 50\;\mu{\rm eV}$, while
the dipole magnets used in ``helioscope''\cite{tokio} experiments are
not sensitive to solar axions heavier than $m_a\simeq 0.03\;{\rm
eV}$.

A pilot experiment carried out by the SOLAX Collaboration\cite{cosmesur}
has already searched for axion
Primakov conversion in a germanium crystal of 1 kg obtaining the
limit $\gagamma\lsim 2.7 \times 10^{-9}\; {\rm GeV}^{-1}$.
This is the (mass independent but solar model dependent) most
stringent laboratory bound for the axion--photon coupling obtained
so far, although less restrictive than the globular cluster
bound\cite{hb} $\gagamma\lsim 0.6 \times 10^{-10}\; {\rm GeV}^{-1}$.
Notice however that the experimental
accuracy of solar observations is orders of
magnitude better than for any other star.

Nevertheless the solar model itself already requires\cite{helio}
$\gagamma\lsim 10^{-9}$ ${\rm GeV}^{-1}$, whereas the above Ge
crystal bound has not yet reached such sensitivity.
The $10^{-9} {\rm GeV^{-1}}$ limit
sets a minimal goal for the sensitivity of future
experiments, prompting the need for a systematic discussion
of present efforts and future prospects for axion searches with crystals.
In the following we give the result of such an analysis,
focusing on Germanium, ${\rm TeO}_2$ and NaI detectors.
%popular detectors (Germanium, ${\rm TeO}_2$ and
%NaI).
%It has recently been stressed\cite{helio} that the model yielding
%the solar axion fluxes used to calculate the expected signals is
%non compatible with the constraints coming from helioseismology if
%$\gagamma \gsim 10^{-9} {\rm GeV}^{-1}$. This would
%imply a possible inconsistency for solar axion limits above that
%value, and sets a minimal goal for the sensitivity of future
%experiments.

\section{Primakov conversion in crystals}

We will make use of the calculation of the flux of solar axions
of Ref. \cite{raffelt} with the modifications introduced in Ref.
\cite{creswick} to include helium and metal diffusion in the solar
model.
% with no direct coupling of the
%axion with electrons.
%so our results will refer to
%the so called {\it hadronic axion}.
%Axions can be efficiently produced in the interior of the Sun
%by Primakov conversion of the blackbody photons in the fluctuating electric field
%of the plasma.
%The resulting flux has an outgoing average axion energy
%$E_a$ of about 4 keV, which is of the order of
%the temperature in the core of the Sun, $T\sim 10^7 K^0\sim 1\;
%{\rm keV}$.
A useful parametrization of the flux is the following:
\begin{equation}
\frac{d\Phi}{dE_a}=\sqrt{\lambda}\frac{\Phi_0}{E_0}\frac{(E_a/E_0)^3}{e^{E_a/E_0}-1}
\label{eq:flux}
\end{equation}
\noindent where
$\lambda$=($\gagamma\times 10^8/$${\rm GeV}$$^{-1}$)$^4$
is an adimensional coupling introduced for later convenience,
$\Phi_0$=5.95 $\times$$10^{14}$ cm$^{-2}$
sec$^{-1}$ and $E_0$=1.103 keV.

In the general case of a multi--target crystal, we calculate
the expected axion--to--photon conversion count rate in a solid--state detector,
integrated in the (electron--equivalent) energy window
$E_1$$<$$E_{ee}$$<$$E_2$, which is given by:
\begin{eqnarray}
R(E_1,E_2)&=&(2\pi)^3 2 \hbar c \frac{V}{v_a^2}\sum_G\frac{d \Phi}{d E_a}
\frac{1}{|\vec{G}|^2}\frac{\gagamma^2}{16 \pi^2}\times
|\sum_j F_{a,j}^0(\vec{G})S_j(\vec{G})|^2\cdot\nonumber\\
&&\sin^2(2\theta)\frac{1}{2}
\left[ {\rm erf}\left( \frac{E_a-E_1}{\sqrt{2}\sigma}\right)-
{\rm erf}\left(\frac{E_a-E_2}{\sqrt{2}\sigma} \right) \right]
\label{eq:final2}
\end{eqnarray}
where we have used the cross--section of the conversion process calculated
in Ref.\cite{cross}.
The first sum is over the vectors
$\vec{G}$ of the reciprocal lattice,  defined by the
property $\exp{i \vec{G}_i\cdot \vec{x}_i}\equiv 1$, where $x_i$
indicate the positions in space of the target nuclei.
$V$ is the volume of the detector,
$v_a$ that of the elementary cell,
$2 \theta$ the scattering angle,
$\sigma$ the resolution of the detector, FWHM=2.35 $\sigma$,
while:
\begin{equation}
S_j(\vec{G})=\sum_i e^{i \vec{a}_i^j \vec{G}}
\nonumber
\end{equation}
\noindent is the structure function of the crystal and
\begin{equation}
F^0_{a,j}(\vec{q})=\frac{Z_j e k^2}{r_j^{-2}+q^2}.
\end{equation}
$k$$\equiv$$|\vec{k}|$$\simeq$$E_a$ is the axion momentum.
The crystal is described by
a Bragg lattice with a basis whose sites are occupied
by atoms of different types. The $\vec{a}_i^j$ indicate the $i$'th basis vector
occupied by the $j$'th target--nucleus type, $Z_j$ is the atomic number of the
$j$--th type target nucleus while
$r_j$$\simeq$1 ${\rm \AA}$ is
the screening length of the corresponding atomic electric field parametrized
with a Yukawa--type potential.

The energy distribution of Eq.(\ref{eq:flux}) implies that the transferred momentum
$q\equiv|\vec{q}|=2 k \sin\theta$
corresponds to a wavelength of a few ${\rm \AA}$, which is of the
order of the distances between atoms in a typical crystal lattice.
This is the reason why a Bragg--reflection pattern arises in the
calculation and in Eq.(\ref{eq:final2}) the integral over the transferred momentum
has been replaced by a sum over the vectors of the reciprocal lattice,
i.e. over the peaks that are produced when the Primakov conversion verifies the
Bragg condition $\vec{q}=\vec{G}$ and the crystal interacts in a coherent way.
The Bragg condition implies that in Eq. (\ref{eq:final2})
$E_a$$=$$\hbar c |\vec{G}|^2/2$
$\hat{u}\cdot \vec{G}$ where the unitary vector $\hat{u}$ points toward
the Sun. This term induces a time dependence in the expected signal
as the detector moves daily around the Sun.

\section{Time correlation and background rejection}
\begin{figure}[t]
\begin{center}
\mbox{\psfig{figure=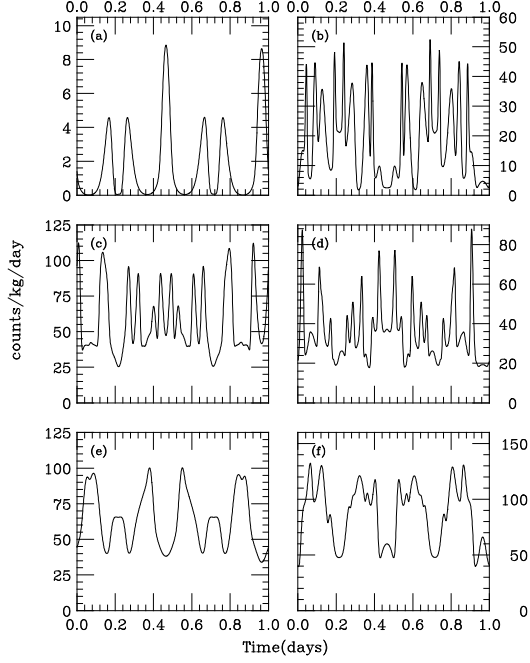,height=90mm}}
\caption{Expected axion signals for Primakov conversion in various
crystals as a function of time for $\lambda=1$.
In the calculation the representative
day of 1 april 1998 and the coordinates of the LNGS laboratory
have been assumed. From top--left to bottom--right:
a) Ge, 2 keV$\leq E_{ee}\leq$2.5 keV;
b) Ge, 4 keV$\leq E_{ee}\leq$4.5 keV; c) TeO$_2$, 5 keV$\leq E_{ee}\leq$7 keV;
d) TeO$_2$, 7 keV$\leq E_{ee}\leq $9 keV; e) NaI, 2 keV$\leq E_{ee}\leq $4 keV;
f) NaI, 4 keV$\leq E_{ee}\leq$6 keV.
\label{fig:axions}}
\end{center}
\end{figure}
In the expected signal the dependence on
$\lambda$ can be factorized: $R\equiv \lambda \bar{R}$.
An example of the function $\bar{R}$ for several materials
is shown in Fig. \ref{fig:axions} as a function of time during
one day. For the crystallographic inputs in the calculation, see for instance
Refs.\cite{cristalografia,cristalteo2}.

The signal is peaked around the maximum of the flux of
Eq.(\ref{eq:flux})
and presents a strong sub--diary
dependence on time, due to the motion of the Sun in the sky.
The time duration of the peaks decreases with growing energies,
from tens of minutes in the lowest part of the axion energy
window, down to a minimum of about one minute in the higher one, and is related to
the energy resolution of the detector.

In order to extract the signal from the background for each energy interval
$E_k<E<E_k+\Delta E$
we introduce, following Ref.\cite{cosmesur}, the quantity:
\begin{equation}
\chi=\sum_{i=1}^{n} \left[ \bar{R}(t_i)-<\bar{R}>\right]\cdot
n_i\equiv \sum_{i=1}^{n} W_{i} \cdot
n_{i}
\label{eq:chi}
\end{equation}
\noindent where the
$n_i$ indicate the number of measured events in the time bin
$t_i,t_i+\Delta t$ and the sum is over
the total period $T$ of data taking. The brackets indicate time average.

By definition the quantity $\chi$
is expected to be compatible with zero in absence of a signal,
while it weights positively the events
recorded in coincidence with the expected peaks.

The time distribution of $n_i$ is supposed to be Poissonian:
\begin{equation}
<n_i>=\left[ \lambda \bar{R}(t_i)+b \right]\Delta t.
\label{eq:mediaconti}
\end{equation}
Assuming that the background $b$ dominates over the signal
the expected average
and variance of $\chi$ are given by:
\begin{eqnarray}
<\chi>&=&\lambda \cdot A\\
\sigma^2(\chi)&\simeq& b/A
\label{eq:medie}
\end{eqnarray}
\noindent with $A\equiv \sum_i W_i^2\Delta t$.
Each energy bin $E_k,E_k+\Delta E$ with background $b_k$
provides an independent estimate $\lambda_k=\chi_k/A_k$
so that one can get the most probable combined value of $\lambda$:
\begin{eqnarray}
\lambda&=&\sum_k\chi_k/\sum_k A_k \nonumber \\
\sigma(\lambda)&=&\left(\sum_k A_k/b_k\right)^{-\frac{1}{2}}.
\label{eq:lambda}
\end{eqnarray}

The sensitivity of an axion experimental search can be expressed
as the upper bound of $\gagamma$ which such experiment would
provide from the non--appearance of the axion signal, for a given
crystal, background and exposure.
If $\lambda$ is compatible to zero, then at the 95\% C.L.
$\lambda\lsim$ 2$\times$ 1.64$\times\sigma(\lambda)$. It is easy
to verify that the ensuing limit on the axion--photon coupling
$\gagamma^{lim}$ scales with the background and
exposure in the following way:
\begin{equation}
%\gagamma\leq \gagamma^{lim}\simeq K \left(\frac{b/{\rm cpd/kg/keV}}{M/{\rm kg}\times
%T/{\rm years}} \right)^{\frac{1}{8}}\times 10^{-9} \; {\rm GeV}^{-1}
\gagamma\leq \gagamma^{lim}\simeq K \left(\frac{\rm b}{{\rm cpd/kg/keV}}\times\frac{\rm kg}
{{\rm M}}\times\frac{\rm years}{\rm T} \right)^{\frac{1}{8}}\times 10^{-9} \; {\rm GeV}^{-1}
\label{eq:limit}
\end{equation}
\noindent where $M$ is the total mass
and $b$ is the average background.
The factor $K$ depends on the parameters of the
crystal, as well as on the experimental threshold and resolution.

The application of the statistical analysis described above
results in a background rejection of about two orders of
magnitude.
In Table \ref{tab:exp} the result of the experiment of Ref.\cite{cosmesur}
is compared to the limits attainable with
running\cite{dama},
being installed\cite{cuoricino,nai_canfranc} and
planned\protect\cite{cuore,genius}
crystal detector experiments.

% of some of the next-generation dark matter
%searches that are being planned or proposed.
%Some examples of attainable limits are given in Table \ref{tab:exp}
%where the sensitivity of the experiment of Ref.\cite{cosmesur} is
%compared to that of some of the next-generation dark matter
%5searches that are being planned or proposed.
\begin{table}
\caption{
Axion search sensitivities for running (DAMA\protect\cite{dama}),
being installed (CUORICINO\protect\cite{cuoricino}, CANFRANC\protect\cite{nai_canfranc})
and planned (CUORE\protect\cite{cuore}, GENIUS\protect\cite{genius}) experiments
are compared to the result of SOLAX\protect\cite{cosmesur}.
The coefficient $K$ is defined in Eq.(\ref{eq:limit}).
\label{tab:exp}
}
\vspace{0.4cm}
\begin{center}
\begin{tabular}{|c|c|c|c|c|c|c|c|}
\hline
  & {\bf K} & {\bf M} & {\bf b (cpd/} & {\bf $E_{th}$}&
 {\bf FWHM } & {\bf $\gagamma^{lim}(2\; {\rm years})$} \\
 &         &    {\bf (kg)} & {\bf /kg/keV)} & {\bf (keV)} & {\bf (keV)} &
 {\bf (GeV$^{-1}$)} \\ \hline
{\bf Ge}\cite{cosmesur} & 2.5 & 1 & 3 & 4 & 1 & 2.7$\times
10^{-9}$\\
{\bf Ge}\cite{genius} & 2.5 & 1000 & 1$\times 10^{-4}$ & 4 & 1 & 3$\times
10^{-10}$\\
{\bf TeO$_2$}\cite{cuoricino}& 3 & 42 & 0.1 & 5 & 2 &
1.3$\times 10^{-9}$ \\
{\bf TeO$_2$}\cite{cuore}& 2.8 & 765 & $1\times 10^{-2}$ & 3 & 2 &
6.3$\times 10^{-10}$ \\
{\bf NaI}\cite{dama} & 2.7 & 87 & 1 & 2 & 2 & 1.4$\times 10^{-9}$ \\
{\bf NaI}\cite{nai_canfranc} & 2.8 & 107 & 2 & 4 & 2 & 1.6$\times 10^{-9}$ \\
\hline
\end{tabular}
\end{center}
\end{table}

\section{Discussion and conclusions}

As shown in the expression of the $\gagamma$ bound of Eq.(\ref{eq:limit})
the improvement in background and accumulation of statistics
is washed out by the 1/8 power dependence of $\gagamma$ on such
parameters.
It is evident, then, from Table \ref{tab:exp} that crystals have no realistic chances
to challenge the globular cluster limit.
A discovery of the axion by this technique would presumably imply
either a systematic error in the stellar--count observations in globular
clusters or a substantial change in the theoretical models
that describe the
late--stage evolution of low--metallicity stars.
%$\gagamma\leq 0.6\times
%10^{-10}$ given by Red Giants and HB stars.

On the other hand,
%, if axions fall in a loophole of present stellar evolution models,
the sensitivity required for crystal--detectors in order to explore a
self--consistent range of $\gagamma$,
compatible with the solar limit of Ref.\cite{helio},
appears to be within reach,
provided that large improvements of background as well as
substancial increase of statistics be guaranteed.
Collecting a statistics of the order of a few tons$\times$year could be
not so difficult to achieve by adding properly the results of various experiments.
In such a case the exploration of a particular axion window,
not accessible to detectors of other types,
could be only a question of time,
as a bonus from current and future dark matter searches.

\section*{Acknowledgements}
This search has been partially supported by the Spanish Agency of
Science and Technology (CICYT) under the grant AEN98--0627. S.S. is
supported by a fellowship of the INFN (Italy).
The authors wish to thank Frank Avignone for useful discussions.

\section*{References}
\bibliography{axion}
\end{document}